\documentclass[10pt]{article}

\usepackage{authblk}
\usepackage{natbib}
\usepackage{mathtools}
\usepackage{amsmath}
\usepackage{amsthm}
\usepackage{amssymb}
\usepackage{amsbsy}
\usepackage{amsfonts}
\usepackage{amscd}
\usepackage{mathrsfs}
\usepackage{bm}
\usepackage{breqn}
\usepackage{color, soul}
\usepackage{enumerate}
\usepackage{empheq}
\usepackage{rotating}
\usepackage{ftnxtra}
\usepackage{fnpos}
\usepackage[titletoc,toc,title]{appendix}
\usepackage{euscript}
\usepackage{graphicx}
\usepackage{epsfig}
\usepackage{epstopdf}
\DeclareGraphicsExtensions{.pdf,.png,.jpg,.eps}
\usepackage{pstool}
\usepackage{upgreek}
\usepackage{mathrsfs}

\usepackage{xcolor}

\usepackage{psfrag}

\numberwithin{equation}{section}

\theoremstyle{plain}	
\newtheorem{thm}{Theorem}[section]

\newtheorem{prop}[thm]{Proposition}
\newtheorem*{prop*}{Proposition} 
\theoremstyle{definition}	

\newtheorem{remark}[thm]{Remark}

\usepackage{graphicx}
\usepackage{lipsum}

\usepackage{caption}
\usepackage{subcaption}

\usepackage{hyperref}
\hypersetup{colorlinks=true, linkcolor=blue}
\hypersetup{colorlinks=true,citecolor=blue}

\DeclareMathAlphabet{\mathpzc}{OT1}{pzc}{m}{it}

\usepackage{amsmath, amsthm, amssymb}

\usepackage{cleveref}

\DeclarePairedDelimiter\abs{\lvert}{\rvert}

\makeatletter
\newsavebox{\@brx}
\newcommand{\llangle}[1][]{\savebox{\@brx}{\(\m@th{#1\langle}\)}%
  \mathopen{\copy\@brx\mkern2mu\kern-0.9\wd\@brx\usebox{\@brx}}}
\newcommand{\rrangle}[1][]{\savebox{\@brx}{\(\m@th{#1\rangle}\)}%
  \mathclose{\copy\@brx\mkern2mu\kern-0.9\wd\@brx\usebox{\@brx}}}%
\let\oldabs\abs
\def\abs{\@ifstar{\oldabs}{\oldabs*}}
\makeatother

\usepackage{accents}

\newcommand{\cCs}{\accentset{s}{\mathsf{C}}}
\newcommand{\cCa}{\accentset{a}{\mathsf{C}}}		

    {\end{bmatrix}}%

\usepackage{enumitem}

\usepackage[utf8]{inputenc}
\usepackage{dutchcal}
\usepackage{multirow}

\begin{document}


\title{\textbf{Universal Displacements in \\Anisotropic Linear Cauchy Elasticity}}

\author[1,2]{Arash Yavari\thanks{Corresponding author, e-mail: arash.yavari@ce.gatech.edu}}
\author[3]{Dimitris Sfyris}
\affil[1]{\small \textit{School of Civil and Environmental Engineering, Georgia Institute of Technology, Atlanta, GA 30332, USA}}
\affil[2]{\small \textit{The George W. Woodruff School of Mechanical Engineering, Georgia Institute of Technology, Atlanta, GA 30332, USA}}
\affil[3]{\small \textit{Institute of Applied and Computational Mathematics (IACM), Foundation for Research and Technology (FORTH), Heraklion, Greece}}

\maketitle

\begin{abstract}
Universal displacements are those displacements that can be maintained for any member of a specific class of linear elastic materials in the absence of body forces, solely by applying boundary tractions. For linear hyperelastic (Green elastic) solids, it is known that the space of universal displacements explicitly depends on the symmetry group of the material, and moreover, the larger the symmetry group the larger the set of universal displacements. Linear Cauchy elastic solids, which include linear hyperelastic solids as a special case, do not necessarily have an underlying energy function. Consequently, their elastic constants do not possess the major symmetries. In this paper, we characterize the universal displacements of anisotropic linear Cauchy elasticity. We prove the unexpected result that for each symmetry class, the set of universal displacements of linear Cauchy elasticity is identical to that of linear hyperelasticity.
\end{abstract}

\begin{description}
\item[Keywords:] Universal deformations, universal displacements, linear elasticity, Cauchy elasticity, Green elasticity, hyperelasticity, odd elasticity.
\end{description}

\tableofcontents

\section{Introduction}\label{sec:intro}

Universal deformations for a given class of solids are those deformations that can be maintained in the absence of body forces and by only applying boundary tractions for any member of the class.
Their study was motivated by the seminal papers of Ronald Rivlin who found several exact solutions in nonlinear hyperelasticity \citep{Rivlin1948, Rivlin1949a, Rivlin1949b}. It was Jerry Ericksen who initiated the first systematic studies of universal deformations in homogeneous compressible and incompressible isotropic hyperelastic solids in two seminal papers \citep{Ericksen1954,Ericksen1955}.
For compressible isotropic hyperelastic solids, \citet{Ericksen1955} showed that homogeneous deformations are the only universal deformations. The problem of characterizing universal deformations is much harder for incompressible isotropic hyperelastic solids. 
Through an elegant and remarkable analysis, \citet{Ericksen1954} discovered four families of universal deformations (besides isochoric homogeneous deformations) for incompressible isotropic solids.
In his analysis the case of constant principal invariants of deformation needed to be treated separately. He conjectured that a deformation with constant principal invariants must be homogeneous. This turned out to be incorrect \citep{Fosdick1966} and led to the discovery of a fifth family of universal deformations \citep{SinghPipkin1965,KlingbeilShield1966}. To this day, it is not known if there are any other constant-principal-stress universal deformations for incompressible isotropic hyperelastic solids---\emph{Ericksen's problem}. In the past few decades, there have been many attempts to solve this open problem \citep{Marris1970,Kafadar1972,Marris1975,Marris1982} and some special cases have also been analyzed \citep{Fosdick1969,Fosdick1971}.
It should be mentioned that universal deformations have played a pivotal role in various developments of nonlinear elasticity:
\begin{itemize}[topsep=2pt,noitemsep, leftmargin=10pt]
\item Universal deformations have played a foundational role in the development of semi-inverse solutions in nonlinear elasticity \citep{Knowles1979,Polignone1991,DePascalis2009,Tadmor2012, Goriely2017}, and more recently, in anelasticity \citep{KumarYavari2023} and viscoelasticity \citep{SadikYavari2024}.
\item Universal deformations provide valuable insights for designing experiments aimed at determining the constitutive relations of specific materials \citep{Rivlin1951,DoyleEricksen1956}.
\item These exact solutions have been utilized as benchmark problems in computational mechanics \citep{Chi2015, Shojaei2018}.
\item Universal deformations have been instrumental in deriving effective properties for nonlinear composites \citep{Hashin1985, Lopez2012, Golgoon2021}.
\end{itemize}

The study of universal deformations has been extended to inhomogeneous and anisotropic hyperelastic solids \citep{Ericksen1954Anisotropic,YavariGoriely2021,Yavari2021,YavariGoriely2023}, anelasticity \citep{YavariGoriely2016,Goodbrake2020,YavariPradhan2022,YavariAccretion2023,PradhanYavari2023}, and liquid crystal elastomers \citep{LeeBhattacharya2023,MihaiGoriely2023}.

In linear elasticity, the counterparts of universal deformations are universal displacements \citep{Truesdell1966,Gurtin1972,Yavari2020}.
The first systematic study of universal displacements in anisotropic linear elasticity was conduced in \citep{Yavari2020}. It was shown that the set of universal displacements explicitly depends on the symmetry class; the larger the symmetry group the larger the corresponding set of universal displacements. Thus, isotropic solids have the largest set of universal displacements and triclinic solids have the smallest one (actually, the only universal displacements for triclinic solids are homogeneous displacements).
The study of universal displacements has been extended to  inhomogeneous solids \citep{YavariGoriely2022}, compressible anisotropic linear elastic solids reinforced by a family of inextensible fibers \citet{Yavari2023}, and linear anelasticity \citep{Yavari2022Anelastic-Universality}.

Recently,  \citep{Yavari2024} extended the analysis of universal deformations and inhomogeneities to inhomogeneous compressible and incompressible isotropic Cauchy elasticity. Cauchy elastic solids  may not necessarily have an energy function, and include hyperelastic solids (Green elastic solids) as a special case. 
This suggests that Cauchy elasticity may have more stringent universality and universal inhomogeneity constraints compared to those of hyperelasticity. Consequently, one might expect smaller sets of universal deformations and universal inhomogeneities for Cauchy elasticity relative to those of hyperelasticity. However, it was shown that the universal deformations and inhomogeneities of Cauchy elasticity coincide with those of Green elasticity in both the compressible and incompressible cases.
In this paper we study the analogue of this problem in anisotropic linear Cauchy elasticity.

This paper is organized as follows. In \S\ref{Sec:Linear-Cauchy-elasticity} we briefly review anisotropic linear Cauchy elasticity. Universal displacements of all the eight symmetry classes of linear Cauchy elasticity are characterized in \S\ref{Sec:Universal-displacements}. 
Conclusions are given in \S\ref{Sec:Conclusions}.

\section{Linear Cauchy elasticity} \label{Sec:Linear-Cauchy-elasticity}

In Cauchy elasticity, stress at a specific point and moment in time explicitly depends on the strain at that point and the same moment in time without any history dependence \citep{Cauchy1828, Truesdell1952, TruesdellNoll2004}. However, unlike hyperelasticity (or Green elasticity \citep{Green1838,Green1839,Spencer2015}) an energy function may not always exist.
\citet{GreenNaghdi1971} demonstrated that Cauchy elasticity is consistent with the first and the second laws of thermodynamics. They illustrated that over a closed path in the strain space, the net work in a Cauchy elastic solid may not be zero. Specifically, in the case of a linear stress-strain relationship they showed that the nonzero stress net work is due to the absence of major symmetries.\footnote{This has also been observed in the recent studies on ``odd" elasticity \citep{Scheibner2020, Fruchart2023}.}
There is a lack of consensus in the literature of nonlinear elasticity regarding the viability of Cauchy elasticity. While some researchers dismiss it \citep{Rivlin1986, Casey2005, Carroll2009, Rajagopal2011}, others seem to accept its validity \citep{Truesdell1952,Ericksen1956,TruesdellNoll2004,Ogden1984,Ostrosablin2017,Kruvzik2019, Bordiga2022}.
Our interest in studying Cauchy elasticity stems from its potential to describe the mechanics of active matter. In active matter, with access to external energy sources, the net work over closed loops in strain space may not vanish.
Moreover, the recent interest in the physics literature on ``odd" elasticity \citep{Scheibner2020, Fruchart2023}, which is essentially linearized non-hyperelastic Cauchy elasticity,\footnote{\citet{Ostrosablin2017} uses the terms Cauchy elasticity and \emph{quasielasticity} interchangeably.} is another motivation for studying Cauchy elasticity and its universal displacements for different symmetry classes.

Let us consider a body $\mathcal{B}$ deforming in a Euclidean ambient space $\mathcal{S}$ that is equipped with a Euclidean metric $\mathbf{g}$. The linearized strain at a point $x\in\mathcal{B}$  is defined as
\begin{equation} 
	\boldsymbol{\epsilon}(x)=\frac{1}{2}\left[\nabla\mathbf{u}^\flat(x)
	+(\nabla\mathbf{u}^\flat(x))^{\star}\right]    \,,
\end{equation}
where $\nabla$ is the Levi-Civita connection of $\mathbf{g}$, and $\mathbf{u}$ is the displacement field. In a coordinate chart $\{x^a\}$, displacement has components $u^a$, and $\mathbf{u}^\flat$ has components $u_a=g_{ab}\,u^b$. In coordinates, $\epsilon_{ab}=\frac{1}{2}(u_{a|b}+u_{b|a})$. Obviously, the linearized strain is symmetric by construction, i.e., $\boldsymbol{\epsilon}^\star=\boldsymbol{\epsilon}$ ($\boldsymbol{\epsilon}^\star$ is the dual of $\boldsymbol{\epsilon}$), or in components $\epsilon_{ab}=\epsilon_{ba}$.
In linear Cauchy elasticity the constitutive equations at $x\in\mathcal{B}$ are written as
\begin{equation} 
	\sigma_{ab}(x)=\mathsf{C}_{abcd}(x)\,\epsilon_{cd}(x) \,,\qquad a,b=1,2,3\,,
\end{equation}
where $\boldsymbol{\mathsf{C}}$ is the elasticity tensor and summation over repeated indices is assumed.
The balance of angular momentum $\sigma_{ab}=\sigma_{ba}$ and symmetry of the linearized strain imply the following minor symmetries:
\begin{equation} 
	\mathsf{C}_{abcd}(x)=\mathsf{C}_{bacd}(x)=\mathsf{C}_{abdc}(x)
	 \,.
\end{equation}
However, the major symmetries do not necessarily hold, i.e., in general, $\mathsf{C}_{abcd}(x)\neq \mathsf{C}_{cdab}(x)$. However, one can always write $\mathsf{C}_{abcd}(x)=\cCs_{abcd}(x)+\cCa_{abcd}(x)$, where
\begin{equation} 
	\cCs_{abcd}(x)=\frac{1}{2}\left[\mathsf{C}_{abcd}(x)+\mathsf{C}_{cdab}(x)\right]\,,\qquad
	\cCa_{abcd}(x)=\frac{1}{2}\left[\mathsf{C}_{abcd}(x)-\mathsf{C}_{cdab}(x)\right]\,,
\end{equation}
are the symmetric and anti-symmetric parts of the elasticity tensor, respectively.

There are several works in the literature on characterizing anisotropic elasticity (and viscoelasticity) tensors that lack the major symmetries \citep{RogersPipkin1963,Podio1987,Yong1991,He1996,Ostrosablin2017}.
It is known that three-dimensional isotropic Cauchy elastic solids lack any anti-symmetric elastic constants \citep{RogersPipkin1963,Ostrosablin2017}.\footnote{This has been noted recently in the so-called odd elasticity, which is simply linear non-hyperelastic Cauchy elasticity. Specifically, it has been noted that in $3$D isotropic elasticity there are no ``odd" elastic constants \citep{Scheibner2020,Fruchart2023}.} Consequently, the universal displacements of isotropic linear Cauchy elastic solids are trivially identical to those of isotropic linear hyperelastic solids.
Similarly, cubic Cauchy elastic solids do not possess any anti-symmetric elastic constants either \citep{RogersPipkin1963,Ostrosablin2017}. Consequently, the universal displacements of cubic linear Cauchy elastic solids are trivially identical to those of cubic linear hyperelastic solids.

\section{Universal displacements in  anisotropic linear Cauchy elasticity} \label{Sec:Universal-displacements}

In linear Cauchy elasticity there are eight symmetry classes: triclinic, monoclinic, tetragonal, trigonal, orthotropic, transversely isotropic, cubic, and isotropic \citep{cowin1995anisotropic,Chadwick2001,ting2003generalized,cowin2007tissue,Ostrosablin2017}.

\begin{remark}
At the crystallographic level there are the $32$ crystallographic point groups distributed among $7$ systems \citep{Wondratschek2006,Haussuhl2008,Newnham2005,Zheng1994,ZhengBoehler1994}. To these if one adds the Curie groups (isotropy essentially), one obtains the $8$ symmetry classes of classical linear elasticity as was mentioned above \citep{Chadwick2001}. Within these classes there are the $32$ crystallographic point groups. In classical linear elasticity, material points do not have microstructure and one can think of linear elasticity as a homogenized theory of simple lattices. It is known that simple lattices only realize the holohedral crystal classes (namely the maximal crystallographic point groups $+$ isotropy) \citep{Pitteri1998,Pitteri2002}. This is why in classical linear elasticity one uses only those matrices associated with the holohedral crystallographic point groups. However, for continua with microstructure, e.g., gradient elasticity or micropolar elasticity, this is no longer true as for multilattices (namely Bravais lattices with a basis) all the $32$ crystallographic point groups can be realized \citep{Pitteri1998,Pitteri2002,Sfyris2023,Sfyris2024}. 
\end{remark}

Next, we consider the eight anisotropy classes, excluding the isotropic and cubic classes, and determine their corresponding universal displacements.
Utilizing the bijection $(11,22,33,23,31,12)\rightarrow(1,2,3,4,5,6)$, the constitutive equations in Voigt notation are given by 
\begin{equation} 
	\sigma_{\alpha}=\sum_{\beta=1}^6 C_{\alpha\beta}\,\epsilon_{\beta}
	=\sum_{\beta=1}^6(c_{\alpha\beta}+b_{\alpha\beta})\,\epsilon_{\beta} \,,\qquad 
	\alpha=1,\cdots,6\,,
\end{equation}
where $[c_{\alpha\beta}]$ ($c_{\alpha\beta}=c_{\beta\alpha}$) and $[b_{\alpha\beta}]$ ($b_{\alpha\beta}=-b_{\beta\alpha}$) are, respectively, the symmetric and anti-symmetric $6\times 6$ stiffness matrices.
Thus, one has the following elasticity matrix \citep{Ostrosablin2017}:
\begin{equation} \label{elasticity-matrix}
    \mathbf{c}=\begin{bmatrix}
	c_{11} & c_{12}-b_{12} & c_{13}-b_{13} & c_{14}-b_{14} & c_{15}-b_{15} & c_{16}-b_{16} \\
	c_{12}+b_{12} & c_{22} & c_{23}-b_{23} & c_{24}-b_{24} & c_{25}-b_{25} & c_{26}-b_{26}\\
	c_{13}+b_{13} & c_{23}+b_{23} & c_{33} & c_{34}-b_{34} & c_{35}-b_{35} & c_{36}-b_{36}  \\
	c_{14}+b_{14} & c_{24}+b_{24} & c_{34}+b_{34} & c_{44} & c_{45}-b_{45} & c_{46}-b_{46}  \\
	c_{15}+b_{15} & c_{25}+b_{25} & c_{35}+b_{35} & c_{45}+b_{45} & c_{55} & c_{56}-b_{56}  \\
	c_{16}+b_{16} & c_{26}+b_{26} & c_{36}+b_{36} & c_{46}+b_{46} & c_{56}+b_{56} & c_{66}  
 \end{bmatrix}\,.
\end{equation}
Further constraints can be imposed on the elastic constants by assuming certain symmetries. These constraints span from the extreme case of isotropy, where only $2$ constants are free (no anti-symmetric elastic constants), to the triclinic case where no constraints are imposed, resulting in $36$ free constants ($21$ symmetric and $15$ anti-symmetric).

\subsection{Universal displacements in triclinic linear Cauchy elastic solids} 

For triclinic solids, the identity and its opposite are the only symmetry transformations. Consequently, there are no restrictions, other than positive-definiteness, on the elastic constants in the elasticity matrix. Thus, there are $36$ independent elastic constants in total.
For a body composed of a homogeneous anisotropic linear elastic solid, the balance of linear momentum in Cartesian coordinates $(x_1,x_2,x_3)$ is expressed as 
\begin{equation} 
	\sigma_{ab,b}(x)=\mathsf{C}_{abcd}\,\epsilon_{cd,b}(x)=0 \,,\qquad a=1,2,3\,.
\end{equation}
Using the matrix representation \eqref{elasticity-matrix}, this is rewritten as:
\begin{equation} \label{Universality-Constraints}
    \begin{bmatrix}
    \frac{\partial}{\partial x_1} & 0 & 0 & 0 & \frac{\partial}{\partial x_3} & \frac{\partial}{\partial x_2} \\
    0 & \frac{\partial}{\partial x_2} & 0 & \frac{\partial}{\partial x_3} & 0 & \frac{\partial}{\partial x_1} \\
    0 & 0 & \frac{\partial}{\partial x_3} & \frac{\partial}{\partial x_2}  & \frac{\partial}{\partial x_1} & 0
    \end{bmatrix}
    \begin{bmatrix}
    C_{11} & C_{12} & C_{13} & C_{14} & C_{15} & C_{16} \\
    C_{21} & C_{22} & C_{23} & C_{24} & C_{25} & C_{26}  \\
    C_{31} & C_{32} & C_{33} & C_{34} & C_{35} & C_{36}  \\
    C_{41} & C_{42} & C_{43} & C_{44} & C_{45} & C_{46}  \\
    C_{51} & C_{52} & C_{53} & C_{45} & C_{55} & C_{56}  \\
    C_{61} & C_{62} & C_{63} & C_{46} & C_{56} & C_{66}  
    \end{bmatrix}
    \begin{bmatrix}
    \frac{\partial u_1}{\partial x_1}  \\
    \frac{\partial u_2}{\partial x_2}  \\
    \frac{\partial u_3}{\partial x_3}  \\
    \frac{\partial u_2}{\partial x_3}+\frac{\partial u_3}{\partial x_2}  \\
    \frac{\partial u_1}{\partial x_3}+\frac{\partial u_3}{\partial x_1}  \\
    \frac{\partial u_1}{\partial x_2}+\frac{\partial u_2}{\partial x_1}   
    \end{bmatrix}
    =   \begin{bmatrix}
    0 \\
    0  \\
    0 
    \end{bmatrix}\,.
\end{equation}

The three equilibrium equations above, in the absence of body forces, must hold for arbitrary values of the independent elastic constants. This implies that in each equilibrium equation, the coefficient of each elastic constant must vanish.
This gives a set of PDEs---the \emph{universality constraints}---for the displacement field.
Arbitrariness of the $21$ symmetric elastic constants $c_{\alpha\beta}$ in \eqref{Universality-Constraints} gives the following $18$ PDEs \citep{Yavari2020}
\begin{empheq}[left={\empheqlbrace }]{align} 
	& \frac{\partial^2 u_1}{\partial x_1^2}=\frac{\partial^2 u_1}{\partial x_1\partial x_2}
	=\frac{\partial^2 u_1}{\partial x_1\partial x_3}=0\,, \\
	& \frac{\partial^2 u_2}{\partial x_2^2}=\frac{\partial^2 u_2}{\partial x_2\partial x_1}
	=\frac{\partial^2 u_2}{\partial x_2\partial x_3}=0\,, \\
	& \frac{\partial^2 u_3}{\partial x_3^2}=\frac{\partial^2 u_3}{\partial x_3\partial x_1}
	=\frac{\partial^2 u_3}{\partial x_2\partial x_3}=0\,,\\
	&\frac{\partial^2 u_a}{\partial x_b^2}=0,~~~a\neq b,~ a,b\in\{1,2,3\}\,,\\
	& \frac{\partial^2 u_1}{\partial x_2\partial x_3}
	=\frac{\partial^2 u_2}{\partial x_1\partial x_3}
	=\frac{\partial^2 u_3}{\partial x_1\partial x_2}=0
	\,.
\end{empheq}
\citet{Yavari2020} showed that the above PDEs imply that the displacement field must be homogeneous.

Looking at the structure of \eqref{elasticity-matrix}, it is easy to see that the universality constraints corresponding to $c_{\alpha\beta}$ include those corresponding to $b_{\alpha\beta}$ ($\alpha\neq \beta$). 
The reason is that the symmetric and antisymmetric elastic constants appear as the pairs $c_{\alpha\beta}+b_{\alpha\beta}$ and $c_{\alpha\beta}-b_{\alpha\beta}$, for $\alpha\neq \beta$, in Navier's equations.
This means that arbitrariness of the anti-symmetric elastic constants in Navier's equations do not give us any new universality constraints. 
This can be verified by direct computations as well.\footnote{All the symbolic computations in this paper were performed using Mathematica Version 13.3.0.0, Wolfram Research, Champaign, IL.}
Hence, we have established the following result.

\begin{prop}The sets of universal displacements of triclinic linear Cauchy elastic solids and triclinic linear hyperelastic solids are identical and coincide with the set of all homogeneous displacement fields.
\end{prop}

\subsection{Universal displacements in monoclinic linear Cauchy elastic solids} 

In monoclinic solids, there exists a plane of material symmetry---a reflection symmetry.
We can assume, without loss of generality, that $x_3$-axis is perpendicular to the plane of material symmetry.
A monoclinic solid has $20$ independent elastic constants ($13$ symmetric and $7$ antisymmetric), and in the Cartesian coordinates $(x_1,x_2,x_3)$, the elasticity matrix takes the following form \citep{Ostrosablin2017}:
\begin{equation}
    \mathbf{C}=\begin{bmatrix}
    c_{11} & c_{12}-b_{12} & c_{13}-b_{13} & 0 & 0 & c_{16}-b_{16} \\
    c_{12}+b_{12} & c_{22} & c_{23}-b_{23} & 0 & 0 & c_{26}-b_{26}  \\
    c_{13}+b_{13} & c_{23}+b_{23} & c_{33} & 0 & 0 & c_{36}-b_{36}  \\
    0 & 0 & 0 & c_{44} & c_{45}-b_{45} & 0  \\
    0 & 0 & 0 & c_{45}+b_{45} & c_{55} & 0  \\
    c_{16}+b_{16} & c_{26}+b_{26} & c_{36}+b_{36} & 0 & 0 & c_{66}  
    \end{bmatrix}.
\end{equation}
Arbitrariness of the symmetric elastic constants in Navier's equations gives us the following set of universality constraints \citep{Yavari2020}:
\begin{empheq}[left={\empheqlbrace }]{align} 
	& \frac{\partial^2 u_1}{\partial x_1^2}=\frac{\partial^2 u_1}{\partial x_1\partial x_2}
	=\frac{\partial^2 u_1}{\partial x_1\partial x_3}=0\,, \\
	& \frac{\partial^2 u_2}{\partial x_2^2}=\frac{\partial^2 u_2}{\partial x_2\partial x_1}
	=\frac{\partial^2 u_2}{\partial x_2\partial x_3}=0\,, \\
	& \frac{\partial^2 u_3}{\partial x_3^2}=\frac{\partial^2 u_3}{\partial x_3\partial x_1}
	=\frac{\partial^2 u_3}{\partial x_2\partial x_3}=0\,,\\
	& \frac{\partial^2 u_1}{\partial x_2^2}=\frac{\partial^2 u_1}{\partial x_3^2}=0\,, \\
	& \frac{\partial^2 u_2}{\partial x_1^2}=\frac{\partial^2 u_2}{\partial x_3^2}=0\,, \\
	& \frac{\partial^2 u_3}{\partial x_1^2}=\frac{\partial^2 u_3}{\partial x_2^2}=0\,,\\	
	& \frac{\partial^2 u_1}{\partial x_2\partial x_3}
	+\frac{\partial^2 u_2}{\partial x_1\partial x_3}=0\,, \\
	& \frac{\partial^2 u_3}{\partial x_1\partial x_2}=0\,.	
\end{empheq}
The arbitrariness of the antisymmetric elastic constants in Navier's equations does not yield any new universality constraints. Therefore, we have established the following result.

\begin{prop}The sets of universal displacements of monoclinic linear Cauchy elastic solids and monoclinic linear hyperelastic solids are identical.
When the planes of symmetry are parallel to the $x_1x_2$-plane, the universal displacements consist of a superposition of homogeneous displacement fields and a one-parameter inhomogeneous displacement field given by $(c\,x_2\,x_3,-c\,x_1\,x_3,0)$.
\end{prop}

\subsection{Universal displacements in  tetragonal linear Cauchy elastic solids} 

In a tetragonal solid, there are five planes of symmetry. The normals of four of these planes are coplanar, while the fifth plane is perpendicular to the other four.
One can assume, without loss of generality, that in the Cartesian coordinate system $(x_1,x_2,x_3)$, the normal corresponding to the fifth plane of symmetry aligns with the $x_3$ axis, with its corresponding plane of symmetry parallel to the $x_1x_2$-plane.
The first two planes of symmetry are parallel to the $x_1x_3$ and $x_2x_3$-planes, respectively. The remaining two planes of symmetry are related to those parallel to the $x_1x_3$-plane through $\frac{\pi}{4}$ and $\frac{3\pi}{4}$ rotations about the $x_3$ axis.
A tetragonal solid possesses $7$ independent elastic constants ($6$ symmetric and $1$ anti-symmetric), and in the Cartesian coordinates $(x_1,x_2,x_3)$, the elasticity matrix takes the following form \citep{Ostrosablin2017}:
\begin{equation}
    \mathbf{C}=\begin{bmatrix}
    c_{11} & c_{12} & c_{13}-b_{13} & 0 & 0 & 0 \\
    c_{12} & c_{11} & c_{13}-b_{13} & 0 & 0 & 0  \\
    c_{13}+b_{13} & c_{13}+b_{13} & c_{33} & 0 & 0 & 0  \\
    0 & 0 & 0 & c_{44} & 0 & 0  \\
    0 & 0 & 0 & 0 & c_{44} & 0  \\
    0 & 0 & 0 & 0 & 0 & c_{66}  
    \end{bmatrix}.
\end{equation}
Navier's equations and arbitrariness of the six symmetric elastic constants yield the following universality constraints \citep{Yavari2020}:
\begin{empheq}[left={\empheqlbrace }]{align} 
	& \frac{\partial^2 u_1}{\partial x_1^2}=\frac{\partial^2 u_1}{\partial x_1\partial x_2}=
	\frac{\partial^2 u_1}{\partial x_2^2}=\frac{\partial^2 u_1}{\partial x_3^2}=0\,, \\
	& \frac{\partial^2 u_2}{\partial x_1\partial x_2}=\frac{\partial^2 u_2}{\partial x_2^2}=
	\frac{\partial^2 u_2}{\partial x_1^2}=\frac{\partial^2 u_2}{\partial x_3^2}=0\,, \\
	&  \frac{\partial^2 u_1}{\partial x_1\partial x_3}
	+\frac{\partial^2 u_2}{\partial x_2\partial x_3}=0\,, \\
	& \frac{\partial^2 u_3}{\partial x_1\partial x_3}=
	\frac{\partial^2 u_3}{\partial x_2\partial x_3}=\frac{\partial^2 u_3}{\partial x_3^2}=0\,, \\
	&  \frac{\partial^2 u_3}{\partial x_1^2}+\frac{\partial^2 u_3}{\partial x_2^2}=0\,.
\end{empheq}
The universality constraints associated with $b_{13}$ are identical to those of $c_{13}$. Consequently, we have established the following result.

\begin{prop}
The sets of universal displacements of tetragonal linear Cauchy elastic solids and tetragonal linear hyperelastic solids are identical.
When the tetragonal axis is parallel to the $x_3$-axis in a Cartesian coordinate system $(x_1,x_2,x_3)$, the universal displacements are a superposition of homogeneous displacement fields and the following inhomogeneous displacement field \citep{Yavari2020}: 
\begin{align} 
	& u_1^{\text{inh}}(x_1,x_2,x_3)=c_1x_2x_3+c_2x_1x_3\,,  \\ 
	& u_2^{\text{inh}}(x_1,x_2,x_3)=-c_2x_2x_3+c_3x_1x_3\,,  \\
	& u_3^{\text{inh}}(x_1,x_2,x_3)=g(x_1,x_2)\,,
\end{align}
where $c_1$, $c_2$, and $c_3$ are constants, and $g=g(x_1,x_2)$ is any harmonic function.
\end{prop}

\subsection{Universal displacements in trigonal linear Cauchy elastic solids} 

A trigonal solid has three planes of symmetry, where the normals of these planes lie in the same plane and are related by rotations of $\frac{\pi}{3}$. In essence, two of the planes of symmetry are obtained from the third one through rotations about a fixed axis by $\frac{\pi}{3}$ and $-\frac{\pi}{3}$. In a Cartesian coordinate system $(x_1,x_2,x_3)$, let us assume that the trigonal axis aligns with the $x_3$-axis.
A trigonal solid possesses $8$ independent elastic constants, consisting of $6$ symmetric and $2$ antisymmetric ones. In the Cartesian coordinates $(x_1,x_2,x_3)$, its elasticity matrix is represented as follows \citep{Ostrosablin2017}:
\begin{equation}
    \mathbf{C}=\begin{bmatrix}
    c_{11} & c_{12} & c_{13}-b_{13} & 0 & c_{15}-b_{15} & 0 \\
    c_{12} & c_{11} & c_{13}-b_{13} & 0 & -c_{15}+b_{15} & 0  \\
    c_{13}+b_{13} & c_{13}+b_{13} & c_{33} & 0 & 0 & 0  \\
    0 & 0 & 0 & c_{44} & 0 & -c_{15}-b_{15}  \\
    c_{15}+b_{15} & -c_{15}-b_{15} & 0 & 0 & c_{44} & 0  \\
    0 & 0 & 0 & -c_{15}+b_{15} & 0 & \frac{1}{2}(c_{11}-c_{12})  
    \end{bmatrix}.
\end{equation}


\begin{remark} The elasticity matrices for the trigonal, hexagonal and transversely isotropic cases given in \citep{Ostrosablin2017} are slightly different from from the commonly used ones found in the literature \citep{Chadwick2001}.
The differences are in the $(6, 6)$ component of \citet{Ostrosablin2017}'s matrices for the trigonal and transversally isotropic cases where a factor $\frac{1}{2}$ is missing as well as in the $(4, 6)$ and $(6, 4)$ components in the trigonal case where a factor $\sqrt{2}$ appears. 
These minor differences, which do not affect the outcome of our analysis, stem from the fact that \citet{Ostrosablin2017} uses the Kelvin instead of the Voigt notation. 
The classical Voigt notation uses the six components of the stresses and strain: $(\sigma_{11}, \sigma_{22}, \sigma_{33}, \sigma_{23}, \sigma_{13}, \sigma_{12} )$ and $(\epsilon_{11}, \epsilon_{22}, \epsilon_{33}, \epsilon_{23}, \epsilon_{13}, \epsilon_{12})$, respectively, while in Kelvin notation $(\sigma_{11}, \sigma_{22}, \sigma_{33}, \sqrt{2}\, \sigma_{23}, \sqrt{2}\, \sigma_{13}, \sqrt{2} \,\sigma_{12} )$ and $(\epsilon_{11}, \epsilon_{22}, \epsilon_{33}, \sqrt{2}\, \epsilon_{23}, \sqrt{2} \,\epsilon_{13}, \sqrt{2} \,\epsilon_{12})$ are used for the stresses and the strains, respectively. 
As pointed out by \citet{cowin1995anisotropic}, the Kelvin notation represents the elastic coefficients as a second-order tensor in a six-dimensional space. However, this is not true when Voigt notation is used.   
\end{remark} 

Navier's equations alongside the arbitrariness of the six symmetric elastic constants yields the following universality constraints \citep{Yavari2020}:
\begin{empheq}[left={\empheqlbrace }]{align} 
	& \frac{\partial^2 u_3}{\partial x_1\partial x_3}
	=\frac{\partial^2 u_3}{\partial x_2\partial x_3}
	= \frac{\partial^2 u_3}{\partial x_3^2}=0\,,\\
	& \frac{\partial^2 u_1}{\partial x_1^2}+\frac{\partial^2 u_1}{\partial x_2^2}
	=\frac{\partial^2 u_2}{\partial x_1^2}+\frac{\partial^2 u_2}{\partial x_2^2}
	=\frac{\partial^2 u_3}{\partial x_1^2}+\frac{\partial^2 u_3}{\partial x_2^2}=0\,,\\
	& \frac{\partial^2 u_1}{\partial x_3^2}=\frac{\partial^2 u_2}{\partial x_3^2}=0\,,\\
	&  \frac{\partial^2 u_1}{\partial x_1\partial x_2}
	= \frac{\partial^2 u_2}{\partial x_1^2}\,,\qquad \frac{\partial^2 u_2}{\partial x_1\partial x_2}
	= \frac{\partial^2 u_1}{\partial x_2^2}\,,\\
	&  2\frac{\partial^2 u_1}{\partial x_1\partial x_3}
	-2\frac{\partial^2 u_2}{\partial x_2\partial x_3}+
	\frac{\partial^2 u_3}{\partial x_1^2}-\frac{\partial^2 u_3}{\partial x_2^2}=0\,,\\
	&  \frac{\partial^2 u_1}{\partial x_2\partial x_3}
	+\frac{\partial^2 u_2}{\partial x_1\partial x_3}
	+\frac{\partial^2 u_3}{\partial x_1\partial x_2}=0\,,\\
	& \frac{\partial^2 u_1}{\partial x_1\partial x_3}
	+\frac{\partial^2 u_2}{\partial x_2\partial x_3}=0\,,\\
	& \frac{\partial^2 u_1}{\partial x_1^2}=3\frac{\partial^2 u_1}{\partial x_2^2}\,.
\end{empheq}
Once more, The arbitrariness of the antisymmetric elastic constants in Navier's equations does not lead to any new universality constraints. Thus, we have proved the following result.

\begin{prop}The sets of universal displacements of trigonal linear Cauchy elastic solids and trigonal linear hyperelastic solids are identical.
The universal displacements consist of a superposition of homogeneous displacements and the following five-parameter inhomogeneous displacement fields \citep{Yavari2020}:
\begin{align} 
	& u_1^{\text{inh}}(x_1,x_2,x_3)=a_{123}\,x_1x_2x_3
	+a_{12}\,x_1x_2+a_{13}\,x_1x_3+a_{23}\,x_2x_3\,,\\
	& u_2^{\text{inh}}(x_1,x_2,x_3)=\frac{1}{2}(a_{12}+a_{123}\,x_3)(x_1^2-x_2^2)
	+b_{13}\,x_1x_3-a_{13}\,x_2x_3\,,\\
	& u_3^{\text{inh}}(x_1,x_2,x_3)=-a_{123}\,x_1^2x_2-(a_{23}+b_{13})\,x_1x_2
	+\frac{1}{3}a_{123}\,x_2^3-a_{13}(x_1^2-x_2^2)\,.
\end{align}
\end{prop}

\subsection{Universal displacements in  orthotropic linear Cauchy elastic solids} 

An orthotropic solid possesses three mutually orthogonal symmetry planes. We assume that these coincide with the coordinate planes in a Cartesian coordinates $(x_1,x_2,x_3)$.
An orthotropic solid has $12$ independent elastic constants, consisting of $9$ symmetric and $3$ antisymmetric ones. In the Cartesian coordinates $(x_1,x_2,x_3)$, its elasticity matrix is represented as follows \citep{Ostrosablin2017}:
\begin{equation}
    \mathbf{C}=\begin{bmatrix}
    c_{11} & c_{12}-b_{12} & c_{13}-b_{13} & 0 & 0 & 0 \\
    c_{12}+b_{12} & c_{22} & c_{23}-b_{23} & 0 & 0 & 0  \\
    c_{13}+b_{13} & c_{23}+b_{23} & c_{33} & 0 & 0 & 0  \\
    0 & 0 & 0 & c_{44} & 0 & 0  \\
    0 & 0 & 0 & 0 & c_{55} & 0  \\
    0 & 0 & 0 & 0 & 0 & c_{66}  
    \end{bmatrix}.
\end{equation}
Navier's equations, along with the arbitrariness of the nine symmetric elastic constants, lead to the following universality constraints \citep{Yavari2020}:
\begin{empheq}[left={\empheqlbrace }]{align} 
	& \frac{\partial^2 u_1}{\partial x_1^2}
	=\frac{\partial^2 u_1}{\partial x_1\partial x_2}
	=\frac{\partial^2 u_1}{\partial x_1\partial x_3}=0, \\
	& \frac{\partial^2 u_2}{\partial x_2^2}
	=\frac{\partial^2 u_2}{\partial x_2\partial x_1}
	=\frac{\partial^2 u_2}{\partial x_2\partial x_3}=0, \\
	& \frac{\partial^2 u_3}{\partial x_3^2}
	=\frac{\partial^2 u_3}{\partial x_3\partial x_1}
	=\frac{\partial^2 u_3}{\partial x_2\partial x_3}=0,\\
	& \frac{\partial^2 u_1}{\partial x_2^2}=\frac{\partial^2 u_1}{\partial x_3^2}=0,\\
	& \frac{\partial^2 u_2}{\partial x_1^2}=\frac{\partial^2 u_2}{\partial x_3^2}=0,\\
	& \frac{\partial^2 u_3}{\partial x_1^2}=\frac{\partial^2 u_3}{\partial x_2^2}=0.
\end{empheq}
Once more, the arbitrariness of the antisymmetric elastic constants in Navier's equations does not lead to any new universality constraints. Therefore, we have established the following result.

\begin{prop}
The sets of universal displacements of orthotropic linear Cauchy elastic solids and orthotropic linear hyperelastic solids are identical.
When the planes of symmetry are perpendicular to the coordinate axes in a Cartesian coordinate system $(x_1,x_2,x_3)$, the universal displacements are a superposition of homogeneous displacement fields and the following three-parameter inhomogeneous displacement field: $(a_1\,x_2\,x_3,a_2\,x_1\,x_3,a_3\,x_1\,x_2)$.
\end{prop}

\subsection{Universal displacements in  transversely isotropic linear Cauchy elastic solids} 

A transversely isotropic solid has an axis of symmetry such that planes perpendicular to it are isotropy planes. Let us assume that the axis of transverse isotropy corresponds to the $x_3$-axis in the Cartesian coordinates $(x_1, x_2, x_3)$. A transversely isotropic solid is characterized by $8$ independent elastic constants ($5$ symmetric and $3$ antisymmetric), and in the Cartesian coordinates $(x_1, x_2, x_3)$, its elasticity matrix is represented as follows \citep{RogersPipkin1963,Ostrosablin2017}:
\begin{equation}
    \mathbf{C}=\begin{bmatrix}
    c_{11} & c_{12} & c_{13}-b_{13} & 0 & 0 & -b_{16} \\
    c_{12} & c_{11} & c_{13}-b_{13} & 0 & 0 & b_{16}  \\
    c_{13}+b_{13} & c_{13}+b_{13} & c_{33} & 0 & 0 & 0  \\
    0 & 0 & 0 & c_{44} & -b_{45} & 0  \\
    0 & 0 & 0 & b_{45} & c_{44} & 0  \\
    0 & 0 & 0 & 0 & 0 & \frac{1}{2}(c_{11}-c_{12})  
    \end{bmatrix}.
\end{equation}
Navier's equations, along with the arbitrariness of the five symmetric elastic constants, give us the following universality constraints \citep{Yavari2020}:
\begin{empheq}[left={\empheqlbrace }]{align} 
	& \label{Trans-1} \frac{\partial^2 u_3}{\partial x_1\partial x_3}
	=\frac{\partial^2 u_3}{\partial x_2\partial x_3}
	= \frac{\partial^2 u_3}{\partial x_3^2}=0,\\
	& \label{Trans-2} 
	\frac{\partial^2 u_3}{\partial x_1^2}+\frac{\partial^2 u_3}{\partial x_2^2}=0,\\
	& \label{Trans-3} 
	\frac{\partial^2 u_1}{\partial x_3^2}=\frac{\partial^2 u_2}{\partial x_3^2}=0,\\
	& \label{Trans-4} 
	\frac{\partial^2 u_1}{\partial x_1^2}+\frac{\partial^2 u_1}{\partial x_2^2}=
	\frac{\partial^2 u_2}{\partial x_1^2}+\frac{\partial^2 u_2}{\partial x_2^2}=0, \\
	& \label{Trans-5} 
	\frac{\partial^2 u_1}{\partial x_2^2}=\frac{\partial^2 u_2}{\partial x_1\partial x_2}, \\
	& \label{Trans-6} 
	\frac{\partial^2 u_2}{\partial x_1^2}=\frac{\partial^2 u_1}{\partial x_1\partial x_2}, \\
	& \label{Trans-7} \frac{\partial^2 u_1}{\partial x_1\partial x_3}+
	\frac{\partial^2 u_2}{\partial x_2\partial x_3}=0.
\end{empheq}
Navier's equations, along with the arbitrariness of the three antisymmetric elastic constants, results in eight universality PDEs governing the universal displacements. However, they are not independent of the above PDEs. Therefore, we have proved the following result.

\begin{prop}
The universal displacements in a transversely isotropic linear Cauchy elastic solid, with the isotropy plane parallel to the $x_1x_2$-plane, are identical to those of the corresponding transversely isotropic linear hyperelastic solids.
The universal displacements have the following form \citep{Yavari2020}:
\begin{equation} 
\begin{aligned}
	& u_1(x_1,x_2,x_3)=c_1x_1+c_2x_2+cx_2x_3+x_3h_1(x_1,x_2)+k_1(x_1,x_2)\,,\\
	& u_2(x_1,x_2,x_3)=-c_2x_1+c_1x_2-cx_1x_3+x_3h_2(x_1,x_2)+k_2(x_1,x_2)\,,\\
	& u_3(x_1,x_2,x_3)=c_3x_3+\hat{u}_3(x_1,x_2)\,,
\end{aligned}
\end{equation}
where $\xi(x_2+ix_1)=h_2(x_1,x_2)+ih_1(x_1,x_2)$ and $\eta\left(x_2+ix_1\right)=k_2(x_1,x_2)+ik_1(x_1,x_2)$ are holomorphic, and $\hat{u}_3(x_1,x_2)$ is harmonic.\footnote{There is a typo in \citep[Proposition~3.6]{Yavari2020} as was pointed out in \citep{YavariGoriely2022}.}
\end{prop}

\section{Conclusions} \label{Sec:Conclusions}

In this paper, we characterized the universal displacements of anisotropic linear Cauchy elasticity. Linear Cauchy elasticity and hyperelasticity have the same number of elastic constants in the case of isotropic and cubic solids. Therefore, in these two cases the sets of their universal displacements coincide trivially.
In the remaining six symmetry classes (triclinic, monoclinic, tetragonal, trigonal, orthotropic, and transversely isotropic), linear Cauchy elasticity has more elastic constants than hyperelasticity: $36$ versus $21$ (triclinic), $20$ versus $13$ (monoclinic), $7$ versus $6$ (tetragonal), $8$ versus $6$ (trigonal), $12$ versus $9$ (orthotropic), and $8$ versus $5$ (transversely isotropic). 
For a given symmetry class, universal displacements are those displacements that in the absence of body forces satisfy Navier's equations for arbitrary elastic constants. 
When there are more elastic constants, there are, in principle, more universality constraints, which potentially could result in a smaller set of universal displacements.
This means that for a given symmetry class (excluding isotropic and cubic), one would anticipate a smaller set of universal displacements for Cauchy elasticity in comparison to hyperelasticity, owing to the larger number of elastic constants in Cauchy elasticity.
However, a closer look at Navier's equations, it becomes apparent that, except for transversely isotropic solids, when there is an antisymmetric elastic constant $b_{\alpha\beta}$ ($\alpha\neq \beta$) it appears as the pairs $c_{\alpha\beta}+b_{\alpha\beta}$ and $c_{\alpha\beta}-b_{\alpha\beta}$ in Navier's equations. This implies that the universality constraints corresponding to $b_{\alpha\beta}$ ($\alpha\neq \beta$) is identical to those of $c_{\alpha\beta}$ ($\alpha\neq \beta$). It turns out that this is the case for transversely isotropic solids as well. 
Therefore, we have established the result that the sets of universal displacements of linear Cauchy elasticity and hyperelasticity are identical for all the eight symmetry classes.

\section*{Acknowledgments}

Arash Yavari was supported by NSF -- Grant No. CMMI 1939901.

\bibliographystyle{abbrvnat}
\bibliography{ref}

\end{document}